\documentclass[12pt]{article}
\usepackage{epsf}
\begin{document}
\small
\author{S. M. Troshin, N. E. Tyurin\\
 Institute for High Energy Physics,\\
Protvino, Moscow Region, 142284 Russia }
\normalsize
\title
{\bf{Beyond the black disk limit: from shadow to antishadow scattering mode}}
\maketitle
\begin{abstract}
New mode in the hadron scattering is predicted to appear at
the energies  beyond $\sqrt{s}\simeq 2$ TeV: the antishadow
scattering mode and the experiments at LHC and VLHC 
in  hadronic reactions will be able to reveal it.
The appearance of the antishadow scattering mode
at these energies   is considered on the basis
of unitarity and geometrical notions of hadron interactions.
Connections with the nonperturbative--QCD models are discussed.
\end{abstract}

\section*{Introduction}

One of the most fundamental discoveries in hadron interactions
at high energies is the rise of total cross--sections with energy.
It is accompanied  by the rise of elastic and inelastic cross--sections
as well as of the ratio of elastic to the total cross--section.

For the first time the total cross--section increase was observed in
$K^+p$--interactions at the Serpukhov accelerator in 1970 \cite{serp}
and it was
discovered later  in  $pp$--interactions at CERN ISR \cite{isr}
and at Fermilab \cite{fnal} in other nucleon-- and meson--proton 
interactions. Recent HERA data \cite{gamp}  demonstrated the rising
behavior of the virtual photon -- proton
total cross-sections.  
Since then a great progress in the experimental and theoretical
studies of  hadronic reactions was achieved. Quantum
Chromodynamics appeared as a theory of strong interactions and
gave an explanation for the behavior of the observables in the
hard hadronic reactions, i.e. the reactions with high momentum
transfers.
However, the dynamics of long distance interactions
(soft processes) is rather
far from its  understanding despite  much work has
been done  in this field. The problems  are  directly
related to the problems of confinement and chiral symmetry
 breaking.
 
 The approaches to soft
hadronic processes are widely varied: Regge--type, geometrical or QCD--inspired
models consider  aspects of such processes from the different
points of view and use various ideas on hadron
structure and interaction dynamics.
The major part of the models  consider the global
characteristics of hadron interactions such as $\sigma _{tot}$,
$\sigma _{inel}$ and $\sigma _{diff}$  related to large
distance interaction dynamics  as reflecting  gross
features of hadron structure \cite{bj}, \cite{petr}.
Despite of the difficulties in application of perturbative QCD for the description
of long--distance interactions and their obvious nonperturbative character,
 it is often possible  to
represent  the high--energy amplitude in the various model
approaches as an expansion over a small parameter
which depends on the kinematics of the process, e.g. for the case of
non--increasing total cross--section the general form of the amplitude
is
\[
F(s,t)=s\sum_n [\tau (s)]^n\exp \left[\frac{a(s)t}{n}\right],
\]
where $\tau(s)\sim 1/\ln s $ is a small parameter at $s\rightarrow \infty$.

Since the expansion is not valid for the rising total 
cross--sections it is possible to find another representation for
that case
with the $t$--dependent expansion parameter \cite{expt}:
\[
F(s,t)=s\sum_{m=1}^\infty [\tau (\sqrt{-t})]^m\Phi _m[R(s), \sqrt{-t}],
\qquad t\neq 0,
\]
where
\[
\tau (\sqrt {-t})=\exp\left(-{\sqrt{-t}}/{\mu _0}\right).
\]
and $\Phi _m[R(s), \sqrt{-t}]$ is an oscillating function of transferred momentum.
The  above formulas as well as some other representations may be successfully
used for the phenomenological analysis of the scattering amplitude
at high energies.

Thus, by now the theoretical treatment of soft hadronic reactions
involves substantial piece of phenomenology and uses various model
 approaches.
They are often based on divergent postulates, but their phenomenological
parts are similar. In particular, an amplitude $V(s,t)$ is considered
as an input for the subsequent unitarization procedure:
\[
F(s,t)=\Phi [V(s,t)].
\]
To reproduce the total cross--section rise the input amplitude
$V(s,t)$ is usually considered as a power function of energy.
This function being taken as an amplitude itself violates
 unitarity  in the direct channel.
To obey unitarity in the direct channel an unitarization
procedure should be used.

There are several ways to restore unitarity  of the scattering
matrix. We  consider  two  schemes: based on the
 use of eikonal and  generalized reaction matrix respectively. There are
also combined methods but those are not often used.
As it was mentioned various models for $V(s,t)$
 may be successfully used to provide
 phenomenological description of high energy hadron scattering.
However, in the particular model
approaches the important dynamical aspects of interaction
could be significantly obscured 
due to  large number of free parameters.

In this paper we discuss some general
properties of hadron scattering, the implications
of unitarity and analyticity,
in particular, manifestations of the antishadow
scattering mode and respective model predictions for the 
observables in elastic scattering and diffraction
dissociation. Our main goal is to draw an attention
to the existence of the antishadow scattering mode at
the energies of LHC and VLHC. It might provide a new 
insight into the dynamics of diffraction and head-on hadronic collisions at superhigh energies.

\section{Geometrical Picture}

In the collisions of two high energy particles the de Broglie
wavelength can be short compared to the typical hadronic size
and hence  optical concepts may be used as  useful guidelines.
Thus, the hadron scattering can be considered 
as a collision of two
relativistically contracted objects of finite size.

The relevant mathematical tool for  description of high energy
hadronic scattering is based on the
impact parameter representation for the scattering amplitude.
 In the case of 
spinless particle scattering this representation has the following form:
\begin{equation}
F(s,t)=\frac{s}{\pi ^2}\int _0^\infty bdb f(s,b)J_0(b\sqrt{-t}).
\label{imp}
\end{equation}
Note that for the scattering of particles with non--zero spin
the impact parameter representation for the helicity amplitudes
 has a similar form with substitution $J_0\rightarrow J_{\Delta\lambda}$,
where $\Delta\lambda$ is the net helicity change between the final
and initial states.
The impact parameter representation as it was shown in \cite{islm}
is valid for all physical energies and scattering angles. This representation
 provides simple semiclassical picture of hadron scattering.

It is often assumed, after the Chou--Yang model was proposed, that the
driving mechanism of hadron scattering is due to overlapping of the two
 matter distributions of colliding hadrons. It could be understood 
 by analogy with  Glauber
 theory of nuclear interactions: one assumes that the matter density
comes from the spatial distribution of hadron constituents and
also assumes  a zero--range interaction between those constituents.
 Such contact interaction might result from the effective
QCD, e.g. based on the Nambu--Jona-Lasinio  Lagrangian.

The important role in the geometrical approach belongs to the
notion of the interaction radius.
The general definition of the interaction radius which is in
 agreement with the above geometrical picture was given in \cite{ngu}:
\begin{equation}
R(s)=l_0(s)/k,
\end{equation}
where $k=\sqrt{s}/2$ is the particle momentum in the c.m.s.
The value for $l_0(s)$ is chosen provided the contributions of the
partial amplitudes from the angular momenta $l> l_0(s)$ are
vanishingly small.

 As a  first approximation one can consider the
energy independent interaction intensity and describe 
the elastic scattering amplitude
in terms of the black disk model where it has
  the form:
\begin{equation}
F(s,t)\propto iR^2(s)\frac{J_1(R(s)\sqrt{-t})}{R(s)\sqrt{-t}}.
\end{equation}
Here $R\sim 1 f$ is the interaction radius. The model is consistent
with the observed structure in the differential cross--sections of $pp$-- and
$\bar p p$--scattering  at $t$ close to  1 $(GeV/c)^2$.

In the simplest case,  neglecting the real part and 
spin,  the impact
parameter amplitude $f(s,b)$ can be obtained as an inverse
transformation according to Eq. \ref{imp} with
\[
F(s,t)\propto\sqrt{s\frac{d\sigma}{dt}(s,t)}.
\]
Thus, one can extract information on the geometrical properties
of interaction from the experimental data.
The analysis of the experimental data on high--energy diffractive scattering
 shows that the effective interaction area expands with energy and the
interaction intensity --- opacity --- increases with energy at
fixed impact parameter $b$. Such analysis used to be carried out
every time as the new experimental data become available. For example
analysis of the  data at the ISR energies (the most
precise data set on differential cross--section for wide $t$--range
available for $\sqrt{s}=53$ GeV) shows that one can observe a central
 impact parameter profile with a tail from the higher partial waves and some
suppression (compared to gaussian) of low partial waves.
 The scattering picture at such energies is close to   gray
 disk with smooth edge which is getting  darker in its center with
energy.

Beside the above simple geometrical observations it is useful
to keep in mind  the
rigorous bounds for the experimental observables.

\section{Bounds for observables and the experimental data}

Bounds for the observables obtained on the firm ground of
general principles such as unitarity and analyticity
are very important for any phenomenological analysis
of soft interactions. However, there are only  few results
obtained on the basis of the axiomatic field theory.

First of all it is the Froissart--Martin bound that gives the
 upper limit
 for the total cross--section:
\begin{equation}
\sigma _{tot}\leq C\ln^2 s,
\end{equation}
where $C=\pi/m_{\pi}^2$ $(=60$mb) and $m_{\pi}$ is the pion mass.

Saturation of this bound, as it is suggested by the existing
 experimental data, imply the dominance of long--distance dynamics.
It also leads to   number of important consequences for the other
observables. For instance, unitarity leads to the following bound
 for elastic cross--section:
\begin{equation}
\sigma_{el}(s)\geq c \frac{\sigma _{tot}^2(s)}{\ln^2 s}.
\end{equation}

Therefore, when the total cross--section asymptotically 
increases as $\ln^2 s$,
elastic cross--section also must rise like $\ln^2 s$. It is
important to note here that there is no similar bound for the
inelastic
cross--section and as we will see further the absence of such bound
 allows for appearance of the antishadow scattering mode at very high
 energies.

If one considers a more general case when
 $\sigma_{tot}\propto \ln ^\gamma s$, then
at asymptotic energies one should have
\begin{equation}
\frac{Re F(s,0)}{Im F(s,0)}\simeq \frac{\gamma\pi}{2\ln s}
\end{equation}
and
\begin{equation}
\frac{\sigma^{\bar a}_{tot}(s)-\sigma^{a}_{tot}(s)}
{\sigma^{\bar a}_{tot}(s)+\sigma^{a}_{tot}(s)}\leq
\ln^{-\gamma/2}(s)
\end{equation}
where $\sigma^{\bar a}_{tot}(s)$ and $\sigma^{a}_{tot}(s)$ are the
total cross--sections of the processes $\bar a +b\rightarrow X$ and
 $a +b\rightarrow X$ correspondingly. In the case of $\gamma =2$ the
total cross--section difference of antiparticle and particle interactions
 should obey the following inequality
\begin{equation}
\Delta\sigma_{tot}(s)\leq\ln s.
\end{equation}

Contrary to the total cross--section behavior, 
the existing experimental data
seem to prefer decreasing $\Delta\sigma_{tot}(s)$. Possible deviations
from such behavior could be expected on the basis
of perturbative QCD \cite{lipa}
 and it was one of the reasons for the recent discussions
 on the Pomeron counterpart --- the Odderon. However, the recent measurements
 real to imaginary part ratio for forward $\bar p p$ scattering
 provide little support for the Odderon. We will not discuss
more thoroughly the interesting problem of
$Re F/Im F$ ratio and  will  consider
for simplicity the case of pure imaginary amplitude.

 For the slope of diffraction cone at $t=0$ in the case of a pure imaginary scattering amplitude the following inequality
takes place:
\begin{equation}
B(s)\geq\frac{\sigma^2_{tot}(s)}{18\pi\sigma_{el}(s)}
\end{equation}
which means that when the total cross--section increases
as $\ln^2 s$, the same dependence is mandatory for the
slope of diffraction cone. It is the stronger shrinkage than
the Regge model predicts: $B(s)\sim \alpha '\ln s$.

 There is also bound \cite{pmpl} for the total cross--section
of single diffractive processes. It was obtained by Pumplin 
in  approach
where inelastic diffraction as well as elastic scattering are assumed
to arise in form of a shadow of inelastic processes and has the
form
\begin{equation}
\sigma_{diff}(s,b)\leq\frac{1}{2}\sigma_{tot}(s,b)-\sigma_{el}(s,b).
\label{pmb}
\end{equation}
The most significant assumption was  that the diffractive eigenamplitudes
in the Good--Walker \cite{gwlk} picture do not exceed the black disk
limit.

At this point some details of the experimental situation have to be
mentioned. At the highest energies the experimental data for the total and elastic
 cross--sections, slope parameter of diffraction cone and cross--section of
 single inelastic diffraction
dissociation have been obtained in $\bar p p$--collisions at
Fermilab. In particular, those measurements show that

\begin{itemize}
\item
the rise of the total cross--section of $p\bar p$--interactions 
is consistent with $\ln^2 s$--dependence, however other dependencies
are not ruled out;
\item
elastic cross--section rises faster that the inelastic and
total cross-sections and has a magnitude  about 1/4 of the total cross-section. 

\end{itemize}
Comparing the value of the elastic to total cross-section ratio
with the lower energy data one can conclude
that the higher the energy, the higher both absolute and
relative probabilities of elastic collisions.

Impact parameter analysis \cite{impan} of the data shows that the scattering amplitude
is probably beyond the black disk limit $|f(s,b)|=1/2$ in head-on collisions.
The Pumplin bound (Eq. \ref{pmb}) is also violated in such collisions and
this is not surprising if one recollects the original ad hoc assumption on the
shadow scattering mode.

\section{Antishadow scattering mode}

The basic role in our consideration belongs to unitarity
of the scattering matrix $SS^+=1$
which is a  reformulation of the probability conservation.
In the impact parameter representation the unitarity equation
rewritten for the elastic scattering amplitude $f(s,b)$ 
at high energies has the form
\begin{equation}
Im f(s,b)=|f(s,b)|^2+\eta(s,b) \label{unt}
\end{equation}
where the inelastic overlap function $\eta(s,b)$ is the sum of
all inelastic channel contributions.  It can be expressed as
a sum of $n$--particle production cross--sections at the 
given impact parameter
\begin{equation}
\eta(s,b)=\sum_n\sigma_n(s,b).
\end{equation}
As it was  mentioned assumption  of a pure
 imaginary amplitude is a rather common approximation at high energies
 and is adequate for our qualitative analysis.
Then the unitarity Eq. \ref{unt} points out that the elastic scattering
amplitude at given impact parameter value
is determined by the inelastic processes.
Eq. \ref{unt} imply the constraint
\[
|f(s,b)|\leq 1
\]
 while the black disk limit
 presumes inequality
\[
|f(s,b)|\leq 1/2.
\]
 The equality $|f(s,b)|=1/2$ corresponds
to maximal absorption in the partial wave with angular momentum
$l\simeq b\sqrt{s}/2$.

The maximal absorption limit is chosen a priori in the eikonal method
of unitarization when the scattering amplitude is written in the
 form:
\begin{equation}
f(s,b)=\frac{i}{2}(1-\exp[i\omega(s,b)])
\end{equation}
and imaginary eikonal $\omega(s,b)=i\Omega(s,b)$ is considered.
The function $\Omega(s,b)$ is called opacity. Eikonal unitarization
 automatically satisfies the unitarity  Eq. \ref{unt} and in
the case of pure imaginary eikonal leads to amplitude which is always
obey the black disk limit.

However, unitarity equation  has the
two solutions for the case of pure imaginary amplitude:
\begin{equation}
f(s,b)=\frac{i}{2}[1\pm \sqrt{1-4\eta(s,b)}].\label{usol}
\end{equation}
Eikonal unitarization with pure imaginary eikonal corresponds to the
choice of the particular
solution with sign minus.

Several models have been proposed for the
eikonal function. For instance, Regge--type models lead to the gaussian
dependence of $\Omega(s,b)$ on impact parameter.  To provide rising
total cross--sections opacity should have a power dependence on energy
\begin{equation}
\Omega(s,b)\propto s^\Delta\exp[-b^2/a(s)],
\end{equation}
where $a(s)\sim \ln s$. In the framework of perturbative QCD--based models
the driving contribution to the opacity is due to jet production in
 gluon--gluon interactions, when
\begin{equation}
\Omega(s,b)\propto \sigma_{jet}\exp[-\mu b],
\end{equation}
where $\sigma_{jet}\sim (s/s_0)^\Delta$.
Such parameterizations lead to the rising total and elastic
 cross--sections and slope parameter:
\begin{equation}
\sigma_{tot}(s)\sim \sigma_{el}(s)\sim B(s)\sim \ln^2 s
\end{equation}
and the ratio
\begin{equation}
\frac{\sigma_{el}(s)}{\sigma_{tot}(s)}\rightarrow \frac{1}{2}.
\end{equation}

 To include the mode where the scattering amplitude exceeds the
 black disk limit one should consider  the
eikonal functions with non--zero
 real parts.  To ensure the  transition from shadow to
 antishadow mode the real part
of eikonal should gain an abrupt increase
equal to $\pi$ at some $s=s_0$. The conventional models do
not foresee such a critical behavior for real part of 
the eikonal.

However, it does not
 mean that the eikonal model itself is in trouble. In particular, the
account for  fluctuations of the eikonal \cite{bars} strongly
 modifies the structure
 of the amplitude and reduces it to algebraic form which is similar to
 that used in the unitarization scheme based  on the
 generalized reaction
matrix.

The latter method
 is based on the relativistic generalization
 of the Heitler equation of radiation dumping\cite{logn}.
 In this approach the elastic scattering amplitude satisfies
unitarity equation since it is constructed as a solution  of  the
following equation \cite{logn} \begin{equation} F = U + iUDF
\label{xx} \end{equation}  presented here in the operator form.
The Eq.\ref{xx} allows  one to satisfy unitarity provided the
 inequality \begin{equation} \mbox{Im} U(s,b) \geq 0 \end{equation}
is fulfilled.
The form of the amplitude in the impact parameter representation
is the following:
\begin{equation}
f(s,b)=\frac{U(s,b)}{1-iU(s,b)}, \label{um}
\end{equation}
where $U(s,b)$ is the generalized reaction matrix, which is considered as an
input dynamical quantity similar to eikonal function.
Similar form for the scattering amplitude was obtained by Feynman in his
 parton model of diffractive scattering \cite{ravn}.
Inelastic overlap function
is connected with $U(s,b)$ by the relation
\begin{equation}
\eta(s,b)=\frac{Im U(s,b)}{|1-iU(s,b)|^{2}}\label{uf}.
\end{equation}

 Construction of  particular models in the framework of the $U$--matrix
approach proceeds with  the same steps as it does
for the eikonal function, i.e. the basic dynamics as well as
the notions on hadron structure
are used  to obtain a particular form for the $U$--matrix.
For example, the Regge--pole approach
\cite{tukh} provides the following form for the $U$--matrix:
\begin{equation}
U(s,b)\propto is^\Delta\exp [-b^2/a(s)],\quad a(s)\sim \alpha '\ln s,
\label{eb2}
\end{equation}
while the chiral quark model which will be discussed below
gives the exponential
$b$--dependence
\begin{equation}
U(s,b)\propto is^\Delta\exp [-\mu b],\label{eb1}
\end{equation}
where $\mu$ is the constant proportional to the masses of the constituent
 quarks.  We have mentioned here only the gross features of those model
 parameterizations without going into the details.

The both parameterizations lead to $\ln^2 s$ rise of the total and
elastic cross--sections and slope parameter $B(s)$:
\begin{equation}
\sigma_{tot}(s)\sim\sigma_{el}(s)\sim B(s) \sim \ln^2 s
\end{equation}
at $s\rightarrow\infty$.
The above results are similar to conclusions of eikonal unitarization.

However, these two unitarization schemes lead to 
different predictions for
 the inelastic cross--sections and for the ratio of elastic to total
cross-section. This ratio in the $U$--matrix unitarization scheme
reaches its maximal possible value at $s\rightarrow \infty$, i.e.
\begin{equation}
\frac{\sigma_{el}(s)}{\sigma_{tot}(s)}\rightarrow 1,
\end{equation}
which reflects in fact that the bound for the partial--wave
 amplitude in the $U$--matrix
approach is $|f|\leq 1$
while  the bound for the case of imaginary eikonal
is (black disk limit):
$|f|\leq 1/2$.

When the amplitude exceeds the black disk limit (in central
collisions at high energies) then the scattering at such impact
parameters turns out to be of an  antishadow nature. It
corresponds to the solution of unitarity equation
Eq. \ref{unt} with  plus sign. In this antishadow scattering mode
 the elastic amplitude increases with decrease of the inelastic
 channels contribution.

The shadow scattering mode is  considered usually as  the only possible
one. But the two solutions of the unitarity
 equation have an equal meaning and the antishadow scattering mode could also
appear in central collisions first as the energy becomes higher.
The both scattering modes are realized in a natural way in the
 $U$--matrix approach despite the two modes are described by the two
 different solutions of unitarity Eq. \ref{usol}.

Let us consider the transition to the antishadow scattering mode
 \cite{phl}. With
conventional parameterizations of the $U$--matrix in the form of Eq. \ref{eb2}
or Eq. \ref{eb1} the inelastic overlap function increases with energies
at modest values of $s$. It reaches its maximum value $\eta(s,b=0)=1/4$ at some
energy $s=s_0$ and beyond this energy the  antishadow
scattering mode appears at small values of $b$. The region of energies and
impact parameters corresponding
to the antishadow scattering mode is determined by the conditions
$Im f(s,b)> 1/2$ and $\eta(s,b)< 1/4$.
The quantitative analysis of the experimental data
 \cite{pras} gives the threshold value of energy: $\sqrt{s_0}\simeq 2$ TeV.

Thus, the function $\eta(s,b)$ becomes peripheral when energy is increasing.
At such energies the inelastic overlap function reaches its maximum
 value at $b=R(s)$ where $R(s)$ is the interaction radius.
So, beyond the transition threshold there are two regions in impact
 parameter space: the central region
of antishadow scattering at $b< R(s)$ and the peripheral region
of shadow scattering at $b> R(s)$.
At $b=R(s)$ the maximal absorbtion (black ring) takes place (Fig. 1).
\begin{figure}[h]
 \vspace*{-0.5cm}
 \begin{center}
 \epsfxsize=120  mm  \epsfbox{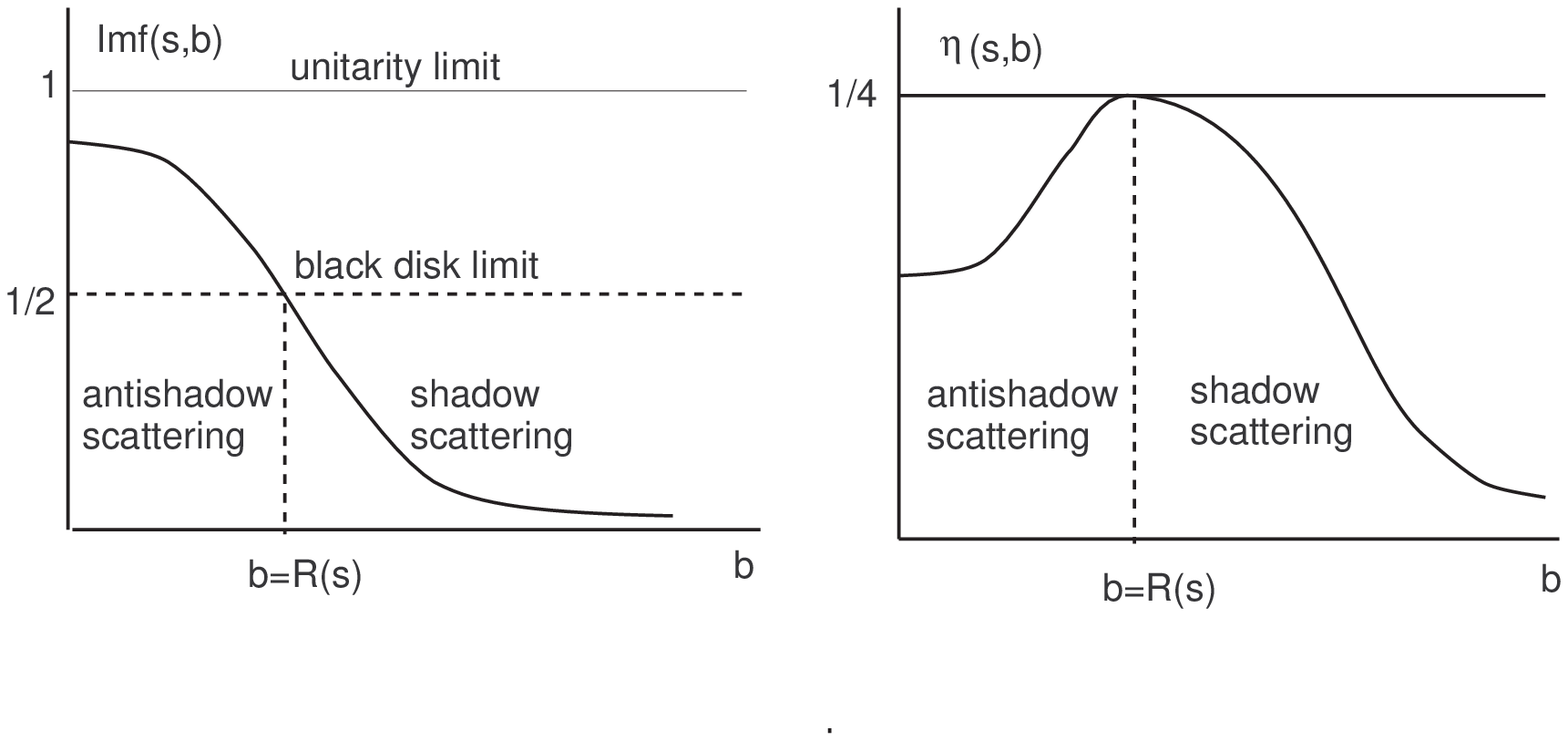}
 \end{center}
 \vspace{-1.5cm}
 \caption{Shadow and antishadow scattering regions}
 \end{figure}

The transition to the antishadow scattering  at small impact
parameters at high energies results also in a relatively slow rise
of inelastic cross--section:
\begin{equation}
\sigma_{inel}(s)=8\pi\int_0^\infty 
\frac{Im U(s,b)}{|1-iU(s,b)|^{2}}\sim\ln s.
\end{equation}
at $s\rightarrow \infty$.

It should be noted that  appearance of the antishadow scattering mode
does not contradict to the basic idea that the particle production
is the driving force for elastic scattering. Indeed, the imaginary
part of the generalized reaction matrix is the sum of inelastic channel
 contributions:
\begin{equation}
Im U(s,b)=\sum_n \bar{U}_n(s,b),\label{vvv}
\end{equation}
where $n$ runs over all inelastic states and
\begin{equation}
\bar{U}_n(s,b)=\int d\Gamma_n |U_n(s,b,\{\xi_n\}|^2
\end{equation}
and $d\Gamma_n$ is the $n$--particle element of the phase space
volume.
The functions $U_n(s,b,\{\xi_n\})$ are determined by the dynamics
 of $2\rightarrow n$ processes. Thus, the quantity $ImU(s,b)$ itself
 is a shadow of the inelastic processes.
However, unitarity leads to  self--damping of the inelastic
channels \cite{bbl} and increase of the function $ImU(s,b)$ results in
decrease
 of the inelastic overlap function $\eta(s,b)$ when $ImU(s,b)$ exceeds unity.

At the energies when the antishadow mode starts to develop (it presumably
could already occur at the energies of the Tevatron--Collider) the Pumplin
 bound Eq. \ref{pmb} for inelastic  diffraction dissociation
cannot be applied since
the main assumption used under its  derivation is not valid any more.

\section{The two modes of hadron scattering and the preasymptotic effects}

 In this section we give a specific analysis of the hadron scattering
 on the basis of particular model.
In Refs. \cite{nuvc,nuov} we   the notions of effective chiral
quark model were used for the description of elastic scattering at small and
large angles.  Hadron dynamics is considered
 in the framework of effective Lagrangian approach.  

A common feature of the chiral models \cite{ball} is the
representation of a baryon as an inner core carrying the baryonic
 charge  and an outer condensate surrounding this core \cite{isl}.
Following these observations it is natural to represent a hadron as
consisting  of  the  inner  region where valence quarks are located
and the outer region filled  with quark condensate \cite{nuov}.  Such
a picture for the hadron structure implies  that  overlapping  and
interaction of peripheral condensates in hadron  collision  occurs
at the first stage. In the overlapping region the condensates
interact and as a result virtual massive quarks appear.  Being
released part  of  hadron  energy  carried  by  the peripheral
condensates goes to a generation of massive quarks. Besides  mass, quark
acquires an internal structure  and a finite size.  Quark radii are
determined by the radii of the clouds.  Strong interaction radius of
quark  $Q$ is determined by its Compton wavelength:  \begin{equation}
r_Q=\xi /m_Q, \label{rq} \end{equation} where constant $\xi$ is
universal for different flavors. 
In the model  valence quarks located in the central part of a hadron
are supposed to scatter in a quasi-inde\-pen\-dent way by the
produced virtual massive quarks at given impact parameter and by  the
 other valence quarks. 

 The function $U(s,b)$  (generalized  reaction matrix)
\cite{logn} --- the basic dynamical quantity of  this approach --- is
chosen as a product of the averaged quark amplitudes \begin{equation}
U(s,b) = \prod^{N}_{Q=1} \langle f_Q(s,b)\rangle \end{equation} in
accordance  with assumed quasi-independent  nature  of  valence
quark scattering.
The $b$--dependence of the function $\langle f_Q \rangle$ related to
 the quark formfactor $F_Q(q)$ has a simple form $\langle
f_Q\rangle\propto\exp(-m_Qb/\xi )$.

Thus, the generalized
reaction matrix (in a pure imaginary case) gets
the following  form
\begin{equation} U(s,b) = ig\left [1+\alpha
\frac{\sqrt{s}}{m_Q}\right]^N \exp(-Mb/\xi ), \label{x}
\end{equation} where $M =\sum^N_{q=1}m_Q$.

At moderate energies $s\ll s_0$ 
(where $\sqrt{s_0}\equiv m_Q/\alpha$)the
function $U(s,b)$ can be represented in the form
\begin{equation} U(s,b) = i{g} \left [1+N\alpha
\frac{\sqrt{s}}{m_Q}\right] \exp(-Mb/\xi ). \label{xl} \end{equation}
 At very high energies $s\gg s_0$ we
could neglect the energy independent term in (\ref{x}) and rewrite
the expression for $U(s,b)$ as  \begin{equation}
U(s,b)=i{g}\left(s/m^2_Q\right)^{N/2}\exp (-Mb/\xi ).
\label{xh} \end{equation}
Calculation of the scattering amplitude is based on the impact
parameter representation  and the analysis of
singularities of $F(s,\beta )$ in complex $\beta $--plane
\cite{expt}.

Besides the energy dependence of these observables we will emphasize
its dependence on geometrical characteristics of non--perturbative
quark interactions.

The total cross--section has the following energy and quark mass
dependencies \begin{equation} \sigma _{tot}(s)=\frac{\pi  \xi
^2}{\langle m_Q \rangle ^2}\Phi (s,N), \label{y} \end{equation} where
$\langle m_Q \rangle=\frac{1}{N}\sum_{Q=1}^N m_Q $ is the mean value
of the constituent quark masses in the colliding hadrons.  The
function $\Phi$ has the following behavior:  \begin{equation} \Phi
(s,N)=\left\{ \begin{array}{cl} \left(8{g}/N^2\right) \left
[1+N\alpha\sqrt{s}/m_Q\right], & s\ll s_0,\\[2ex] \ln ^2 s, & s\gg
s_0.  \end{array} \right. \label{yy} \end{equation} Thus, at
asymptotically high energies the model provides \[ \lim_{s\rightarrow
\infty}\frac{\sigma_{tot}(\bar a b)} {\sigma_{tot}(ab)}=1.  \]

Linear with $\sqrt{s}$  preasymptotic  rise of the total
cross--sections is in agreement with the experimental data  up to
 $\sqrt{s}\sim 0.5 $ TeV \cite{pras}.

The inelastic cross-section can be calculated in the model
explicitely, viz:  \begin{equation} \sigma _{inel}(s)=\frac{8\pi \xi
^2}{N^2 \langle m_Q \rangle ^2} \ln \left[1+{g}(1+\frac{\alpha
\sqrt{s}}{m_Q})^N\right], \label{s} \end{equation} At asymptotically
high energies the inelastic cross--section rise is as follows
\begin{equation} \sigma _{inel}(s)=\frac{4\pi \xi ^2}{N \langle m_Q
\rangle ^2} \ln s \label{ss} \end{equation}

At $s\gg s_0 $ the dependence of the hadron interaction radius $R(s)$
and the ratio $\sigma_{el}/\sigma_{tot}$ on  $\langle
 m_Q \rangle $ is provided by the following equations:  \begin{eqnarray}
 R(s) & = & \frac{\xi }{2 \langle m_Q \rangle} \ln s, \label{rr}\\
\frac{\sigma_{el}(s)}{\sigma _{tot}(s)} & = & 1-\frac{4}{N \ln s}.
 \label{rs} \end{eqnarray} It is important to note here that such a
behavior of the ratio $\sigma_{el}/\sigma_{tot}$ and $\sigma
_{inel}(s)$ results from self--damping of inelastic channels
\cite{bbl} at small impact distances. Numerical estimates \cite{pras}
show that the ratio $\sigma_{el}(s)/\sigma
_{tot}(s)$ bocomes close to the asymptotic value 1  at extremely high energies $\sqrt{s}=500$ TeV.

Thus, unitarization drastically changes the scattering picture:  at
lower energies inelastic channels provide dominant contribution and
scattering amplitude has a shadow origin while at high energies
elastic scattering dominates over inelasic contribution and the
scattering picture corresponds to the antishadow mode.  The
functional $s$--dependencies of observables also differ
significantly.  For example, $s$--dependence of total cross-section
at $s\ll s_0$ is described by a simple linear function of $\sqrt{s}$.
It has been shown that such dependence does not contradict to the
experimental data for hadron total cross--sections  up to
$\sqrt{s}\sim 0.5$ TeV. Such dependence  corresponds to that of the hard
Pomeron with $\Delta=0.5$, however, it was obtained in different
 approach \cite{nuov}.  This is a preasymptotic dependence and it has
nothing to do  with the true asymptotics of the total cross-sections.
In the model such behavior of the hadronic cross--sections reflects
 the energy dependence of number of virtual quarks generated 
 under condensate collisions in the
intermediate transient stage of hadronic interaction.

\section{ Antishadow scattering mode and 
inelastic diffractive processes}
Inelastic diffractive production as well as elastic scattering at
low transferred momenta are the two basic processes which would lead
to understanding of large distance dynamics and hadron structure. 
Concerning
 inelastic diffractive processes this statement can be traced back to
the seminal paper \cite{gwlk} where such processes were
considered as a result of a difference in absorption of various
proton states.
Later on these states have got a parton--like interpretation.
New
 data were obtained for single diffraction production process
\begin{equation}
h_1+h_2\rightarrow h_1+h_2^* \label{bpr}
\end{equation}
when the hadron $h_2$ is excited to the state $h_2^*$ with
invariant mass $M$ and the same quantum numbers. Its subsequent
decay results in the multiparticle final state. The inclusive
differential cross--section shows a simple dependence on the invariant
mass 
\begin{equation}
\frac{d\sigma _{diff}}{dM^2} \propto \frac{1}{M^2}. \label{mdep}
\end{equation}
However, energy dependence of the diffractive production
cross--section $\sigma _{diff}(s)$ is not so evident from the 
data.  This ambiguity is partly due to difficulties in the
experimental definition of the inelastic diffractive
cross--section.

The particular  experimental regularities observed in diffractive production
can be described in the framework of different approaches. $1/M^2$
dependence is naturally described by the triple--pomeron diagrams
in the framework of Regge--model. The proposed in Ref. \cite{donl}
similarity between the Pomeron and photon exchanges allowed to
calculate  diffractive dissociation cross--section in terms of structure
function $\nu W_2$  measured in deep inelastic lepton
scattering.
Several models use optical picture for the description of
diffractive production \cite{flet} but these models in large
extent concern the angular distribution of diffractive
cross--section and $M^2$--dependence is left beyond of their
scope. The attempt to explain $M^2$--dependence in the
framework of optical model considering diffractive dissociation as
a bremsstrahlung where virtual quanta are released from a strong
field was made in Ref. \cite{fass}.

In this section for description of single diffractive
processes we use model approach described in section 4.

  To obtain
the cross-section  of the diffractive  dissociation  process
 we
should single out among the  final  states in Eq. \ref{vvv} those corresponding  to  the
process (\ref{bpr}) .  Let  for  simplicity  consider again  the
case  of  pure imaginary $U$-matrix. Then we can represent
$d\sigma _{diff}/dM^2$
in the following form
\begin{equation}
\frac{d\sigma _{diff}}{dM^2} = 8\pi
\int^{\infty}_{0} bdb
\frac{U_{diff}(s,b,M)}{[1+U(s,b)]^2}\label{y1}
\end{equation}
where expression for $U_{diff}(s,b,M)$  includes  contributions
from all the
final states $|n\rangle_{diff}$  which  results  from  the  decay  of  the
excited  hadron   $h^*_2$   of  mass   $M$:
 $h^*_2\rightarrow |n\rangle _{diff}$.

For consideration of the  diffractive  production  at the quark
level we extend  the  picture  for  hadron  interaction for
elastic scattering, described in section 4. Since the constituent quark is an extended
object there is a non--zero probability of its excitation at the
first stage of hadron collision during the interaction of
peripheral condensates.  Therefore it is natural to assume  that
the origin of diffractive  production process is the excitation of
one of  the valence  quarks
in colliding hadron: $Q\rightarrow Q^*$, its subsequent scattering
and decay into the final state. The  excited constituent
quark is scattered similar to  other
valence quarks in a quasi-independent way. The function
$U_{diff}(s,b,M)$ can be represented then as a product
\begin{equation}
U_{diff}(s,b,M) = \langle f_{Q^*}(s,b,M_{Q^*})\rangle\prod^{N-1}_{Q=1}
\langle f_Q(s,b)\rangle,
\end{equation}
where $M_{Q^*}$ is the mass of excited constituent quark,  which  is
proportional to the mass $M$ of excited hadron $h_2^*$ for large
values of $M$. The last statement presumes the additivity of
constituent quark masses.
 The $b$--dependence
of the amplitude $\langle f_{Q^*} \rangle $ is related to the
formfactor of excited quark whose radius is detemined by its mass
$M_{Q^*}$ ($r_{Q^*} = \xi /M_{Q^*})$. The expression for
$U_{diff}(s,b,M)$ can be rewritten then in the following form:
\begin{equation}
U_{diff}(s,b,M) =g^* U(s,b)\exp[-(M_{Q*}-m_Q)b/\xi ],
\end{equation}
where constant $g^*$ is  proportional to  the
 relative probability
of excitation of the constituent quark.
The value of $g^*$ is a non-zero one, however,
 $g^*<1$ since we expect that the excitation of
 any constituent quark  has
lower probability compared  to probability for this quark to stay
unexcited.  The
excited quark is not stable and its subsequent decay is associated
with the decay of excited hadron  $h_2^*$  into  the
multiparticle
final state $|n\rangle _{diff}$.

The cross-section of diffractive dissociation process is given  by
expression (\ref{y1}) and has the following $s$ and $M^2$ dependence
\begin{equation}
\frac{d\sigma_{diff}}{dM^2}\simeq
\frac{8\pi  g^*\xi ^2}{(M_{Q^*}-m^2_Q)^2}\eta (s,0)\simeq
\frac{8\pi  g^*\xi ^2}{M^2} \eta(s,0)
\end{equation}
Thus,   we   obtained the familiar
$1/M^2$  dependence  of the  diffraction   cross-section   which
is related in this model to the geometrical size  of excited  constituent
quark.

The double  dissociation  processes
\begin{equation}
h_1+h_2\rightarrow h_1^*+h_2^*
\end{equation}
can  be  considered  on the grounds of previous approach to  the  single  diffractive
dissociation. Here one of the
constituent  quarks   in   each of the colliding   hadrons should be
excited.
Cross-section of double diffraction process has similar $M^2$- and
$s$--dependencies  and is to be  suppresed in  comparison with
the  single diffractive cross-section  by an extra factor  $g^*<1$.

The energy dependence of single diffractive cross-section has  the
following form
\begin{equation}
\sigma_{diff}(s) = 8\pi  g^*\xi ^2\eta(s,0)
\int^{M^2_1}_{M^2_0}
\frac{dM^2}{M^2} = 8\pi  g^*\xi ^2\eta(s,0)
\ln\frac{s(1-x_1)}{M^2_0},\label{z}
\end{equation}
where $x_1$ is the lower limit of the relative momentum of hadron
$h_1(x_1\simeq  0.8-0.9)$ which corresponds to the experimental
constraint on diffractive process. Eq. (\ref{z}) shows   that the   total
cross-section of diffractive dissociation has a non-trivial energy
dependence which is determined by the
 contribution of inelastic  channels  into  unitarity equation at
 zero value of impact parameter. The dependence of $\eta (s,0)$ is
determined  by Eq. (\ref{uf}), where expression for $U(s,b)$ is
given by Eq.
(\ref{x}).
At $s\leq s_0$, ($s_0$ is determined by equation
$|U(s_0, 0)|=1$) $\eta(s,0)$ increases with  energy.
This  increase as it  follows  from  Eq. (\ref{x})  and from
the experimental   data \cite{mitt} is rather slow one. However at
$s\geq s_0$, $\eta(s,0)$ reaches its maximum value $\eta  (s,0)=1/4$
and at $s > s_0$, the function $\eta(s,0)$ decreases with energy. At $s\rightarrow\infty$:
\begin{equation}
\sigma _{diff}(s)
 \propto \left(\frac{1}{\sqrt{s}}\right)^N\ln{s}
\end{equation}
since $\eta (s,0) \propto \left({1}/{\sqrt{s}}\right)^N$ in this
limit.

  Thus  at  asymptotical
energies the inelastic diffraction cross section drops to zero.
Decrease of diffractive production cross--section at high energies
($s>s_0$) is due to the  fact  that $\eta  (s,b)$
becomes peripheral at  $s > s_0$  and  the whole  picture  corresponds  to
the antishadow scattering at $b < R(s)$ and to the shadow scattering at
$b>R(s)$ where $R(s)$ is the interaction radius.
The qulitative behavior of $\sigma _{diff}(s)$ is shown
on Fig. 2. 
\begin{figure}[h]
 \vspace*{-0.5cm}
 \begin{center}
 \epsfxsize= 120 mm  \epsfbox{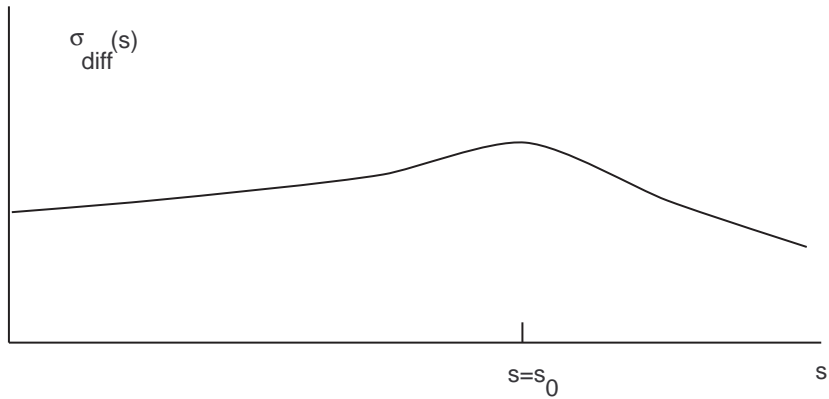}
 \end{center}
 \vspace{-1cm}
 \caption{Energy dependence of diffractive cross-section}
 \end{figure}

 The development of the antishadow mode in
head-on pp and $p\bar p$--collisions  could be associated with new phenomena
 in  the  central  hadronic  collisions  where  the
temperatures are high and the energy density can be up to  several
$GeV/fm^3$. In such collisions the constituent quarks have noticeable
probability to be excited. Due to its high mass and small
transverse size the excited state has low probability of 
interactions
with other particles. It may be also related to an interesting phenomena
 in cosmic ray experiments where particles with abnormal
persistency in lead chambers were observed \cite{albr}.

Of course, there
might be different reasons for  decrease of $\sigma _{diff}(s)$.
The decreasing energy dependence of $\sigma _{diff}(s)$ was also
predicted in Refs. \cite{chg}, \cite{dur}.
 As it
was pointed out in Ref. \cite{gwlk} in the limit of complete
absorption  the diffractive dissociation should vanish.
 It was
advocated in Ref. \cite{bars} that this situation will occur at
superhigh energies and it is the reason for  decrease
 of inelastic diffractive cross--section.
 This is
completely the same behavior as it is predicted by the model
presented, however
in our case the reason for that is the transition to the antishadow
scattering mode in head-on collisions in the  multi--TeV energy range.
It should be noted, however, that the diffractive cross-section at  preasymptotic energies
 has a similar to total and elastic
cross-section energy dependence and it will be discussed in
the concluding part of this paper.

\section{Universal preasymptotics}

The straitforward interpretation of the recent HERA data on the
deep--inelastic scattering together with the analysis of the
 data on hadron--hadron
scattering in  terms of the Regge model could lead to the
unexpected conclusion
on the existence of the  various Pomerons \cite{land} or
the  various manifestations
of  unique Pomeron
in the different processes depending on the typical
scale of the process \cite{brt}.
 The approaches \cite{dl,pr} contending the
dominance of the soft Pomeron do not rule out
 existence of the
hard Pomeron either.

Indeed, soft hadronic reactions imply that the Pomeron's intercept
$\alpha_{\cal{P}}=1.08$ \cite{land}, and small--$x$ dependence of the
structure function $F_2(x,Q^2)$ leads to
$\alpha_{\cal{P}}=1.4$-$1.5$ \cite{h1,zeus} and the measurements of
the diffractive cross--section in the deep inelastic scattering
provide $\alpha_{\cal{P}}=1.23$ \cite{dfr}.  So, does this mean
that  we have few Pomerons or we have few different manifestations
of  the same Pomeron depending on the particular process?  Probably
both options are not to be considered as the firm ones, since the
experimental data  used to advocate these statements were obtained at  not
high enough energies where, in fact, the preasymptotic regime of
interactions does take place.  The above conclusions are based on
the presumed dominance of the Pomeron contribution already in the
preasymptotic energy region. What is called a Pomeron  is to be interpreted as a
true asymptotical contribution of the driving  mechanism.

 In this section we argue that all the three classes of the processes
described above are related to the similar mechanisms and the
corresponding energy dependence of the cross-sections can be well
described by the universal functional energy dependence  of the
type $a+b\sqrt{s}$. Such dependence is valid for the preasymptotic
energy region only and beyond this region unitarity changes the
 picture drastically.  We   consider  for  illustration  the
unitarized chiral quark model (section 4).

Fit to the
total $hp$ cross-sections  gives small values for the parameters
 $g$ and $\alpha$ ( $g,\alpha\ll 1$) \cite{pras}.  It means that at
$s\ll s_0$ the second term in the square brackets in Eqs.
(\ref{um}) and (\ref{uf}) is small and we can expand over it.  The
numerical value of $s_0$ is determined by the equation $|U(s,0)|=1$
and is \cite{pras} $\sqrt{s_0}\simeq 2\,\mbox {TeV}.$ At this energy
the amplitude has the value $|f(s_0,0)|=1/2$.
The value of
$s_0$ is on the verge of the preasymptotic energy region, i.e.  the
Tevatron energy is at the beginning of the road to the asymptotics.
Evidently the HERA energy range $W(=\sqrt{s_{\gamma p}})\leq 300$
GeV is in a preasymptotic domain. 

The above model gives the linear with $\sqrt{s}$ dependence for the
total cross--sections according to Eqs. (\ref{um}) and (\ref{uf}):
\begin{equation} \sigma_{tot}^{hp,\gamma p}=a+b\sqrt{s},\label{lin}
\end{equation} where parameters $a$ and $b$ are different for
different processes and the same is true for the scale $s_0$.  
It was shown \cite{pras} that Eq. \ref{lin}
is in a good agreement with the experimental data.

The same dependence for the total cross--section of $\gamma^* p$
scattering is assumed by the small--$x$ behavior of the structure
function $F_2(x,Q^2)$  \cite{h1,zeus} and
obtained  in  \cite{eps}:  \begin{equation}
F_2(x,Q^2)=a(Q^2)+b(Q^2)/\sqrt{x}.\label{f2} \end{equation} The
experimental data also indicate the critical behavior of the
function
$b(Q^2)$ at $Q^2\simeq 1$ (GeV/c$)^2$. This scale could be related
to the radius of a constituent quark and its  structure.

The third value for the Pomeron intercept $\alpha_{\cal{P}}=1.23$
has been obtained from the analysis of the experimental data on the
diffractive cross--section in deep--inelastic scattering \cite{dfr}
 where the dependence of $d\sigma^{diff}_{\gamma^* p\to XN}/dM^2_X$
on $W$  was parametrized according to the Regge model and the
Pomeron dominance has been assumed:  \begin{equation}
d\sigma^{diff}_{\gamma^* p\to XN}/dM^2_X\propto
(W^2)^{2\alpha_{\cal{P}}-2}.\label{rd} \end{equation}

The data  demontrate  linear rise of the differential
cross--section $d\sigma^{diff}_{\gamma^* p\to XN}/dM^2_X$ with $W$,
i.e.  we observe here just the same functional dependence on the
c.m.s. energy as  for $\sigma_{tot}^{hp,\gamma
p,\gamma^* p}$.  Regarding the preasymptotic nature of the
interaction mode we arrive  to the universal c.m.s. energy
dependence in the framework of the used model.

Indeed, in the framework of this model the hadron inelastic
diffractive cross--section is given by the following expression
\cite{usdf}:  \begin{equation} \frac{d\sigma^{diff}_{hp\to
XN}}{dM_X^2}\simeq \frac{8\pi g^*\xi^2}{M_X^2}\eta (s,0),\label{ds}
\end{equation} where \[ \eta (s,b)=\mbox{Im} U(s,b)/[1-iU(s,b)]^2
\] is the inelastic overlap function.

At the preasymptotic energies $s\ll s_0$ the energy dependence of
inelastic diffractive cross--section resulting from Eq. (\ref{uf})
 is again determined by the generic form \begin{equation}
\frac{d\sigma^{diff}_{hp\to XN}}{dM_X^2} \propto
a+b\sqrt{s}.\label{ds1} \end{equation} Inelastic diffractive
cross--section for the $\gamma^* p$ interactions can be obtained
using for example VMD model, i.e.  \begin{equation}
\frac{d\sigma^{diff}_{\gamma^* p\to XN}}{dM_X^2} \propto
a(Q^2)+b(Q^2)W.\label{ds2} \end{equation} The same functional
dependence can be obtained using the "aligned jet" model \cite{alj}
along with the unitarized chiral quark model \cite{us94}.

 The above linear dependences for the
cross--sections of different processes is the generic feature
associated with the preasymptotic nature of the interaction
dynamics at $s\ll s_0$. As one goes above this  energy range the
function $|U(s,b)|$ is rising and when $|U(s,0)|\geq 1$ the
unitarity starts to play the major role and provides the $\ln^2 s$
rise of the total cross--sections at $s\gg s_0$ \cite{us94} and
also the following behavior of the structure function $F_2(x,Q^2)$
\begin{equation} F_2(x,Q^2)\propto\ln^2( 1/{x}) \end{equation} at
$x\rightarrow 0$ \cite{eps}.  At the same time unitarity leads to
the decreasing dependence of the inelastic diffractive
cross--section at $s\to\infty$ \begin{equation}
\frac{d\sigma^{diff}}{dM_X^2} \propto
\left(\frac{1}{\sqrt{s}}\right)^N.\label{ds3} \end{equation} for
the  $hp$, $\gamma p$ and $\gamma^* p$ processes \cite{usdf}.  Eq. \ref{ds3} is associated with the antishadow scattering mode which
develops  at small impact parameters at $s> s_0$.

Thus, we might expect the different asymptotic and universal
preasymptotic behaviors for the different classes of the
diffraction processes.

To summarize, we would like to emphasize that the unified
description of the processes of $hp$, $\gamma p$ and $\gamma^* p$
diffraction scattering with the universal cross-section dependence
on the c.m.s. interaction energy is possible.  For the illustration
we used the unitarized chiral quark model which has a
nonperturbative origin and leads to the linear c.m.s. energy
dependence of the cross--sections  in the preasymptotic energy
region for the above processes. Universality of such preasymptotic
 behavior agrees with the  experiment.

The assumption on the existence of the different Pomerons results
from the use of the asymptotic formulas in the preasymptotic energy
region and the neglect of the unitarity at higher energies
beyond this preasymptotic region. It should be taken with certain
caution.

\section*{Conclusion}

Studies of soft interactions at the highest energies can
lead to the discoveries of fundamental importance. The genesis of hadron scattering with rising energy  
 can be described as  transition from the grey to black disk and eventually
to black ring with the antishadow scattering mode in the center. 
Such transitions are under control of unitarity
of the scattering matrix.

The appearance of antishadow scattering mode could
be revealed performing impact parameter analysis
of elastic scattering and directly in the measurements
of the inelastic diffractive cross section (cf. Figs. 1,2).

 It would be interesting to speculate on the particular
physical origin of the antishadow scattering mode.  Its
existence  can be correlated with the new phenomena
 expected at high energies in the central hadronic collisions.
 Such collisions are usually associated with the formation of quark--gluon
plasma and disoriented chiral condensate in the inner part of the
interaction region. What are the particular correlations between those phenomena and
the antishadow scattering? The answer  can be obtained in
the  nonperturbative QCD studies  and in the
 experiments devoted to studies of soft processes at LHC
 and VLHC.
  It seems  that the anomalies observed in cosmic ray experiments \cite{albr} might
 also be correlated with  development of the  antishadow scattering mode
 in the  central hadron collisions.

\end{document}